\documentclass[11pt,twoside,a4paper,cr,final]{cms-tdr}
\def\svnVersion{64946M}

\begin{document}

\hyphenation{had-ron-i-za-tion}
\hyphenation{cal-or-i-me-ter}
\hyphenation{de-vices}

\RCS$Revision: 62698 $
\RCS$HeadURL: svn+ssh://tdahms@svn.cern.ch/reps/tdr2/papers/XXX-08-000/trunk/XXX-08-000.tex $
\RCS$Id: XXX-08-000.tex 62698 2011-06-21 00:28:58Z alverson $

\newcommand{\mumu}{\ensuremath{\mu^+\mu^-}\xspace}
\newcommand{\qqbar}{\ensuremath{\text{q}\overline{\text{q}}}\xspace}
\newcommand{\QQbar}{\ensuremath{\text{Q}\overline{\text{Q}}}\xspace}
\newcommand{\Jpsi}{\JPsi}
\newcommand{\psiP}{\ensuremath{\psi\text{(2S)}}\xspace}

\newcommand{\doubleRatioPsi}{\ensuremath{\left.(N_{\psiP}/N_{\Jpsi})_{\PbPb}/(N_{\psiP}/N_{\Jpsi})_{\pp}\right.\xspace}}
\newcommand{\doubleRatioUps}{\ensuremath{\left.(N_{\PgUb}/N_{\PgUa})_{\PbPb}/(N_{\PgUb}/N_{\PgUa})_{\pp}\right.\xspace}}
\newcommand{\doubleRatioUpsC}{\ensuremath{\left.(N_{\PgUc}/N_{\PgUa})_{\PbPb}/(N_{\PgUc}/N_{\PgUa})_{\pp}\right.\xspace}}

\newcommand{\npart}{\ensuremath{N_{\text{part}}}\xspace}
\newcommand{\ncoll}{\ensuremath{N_{\text{coll}}}\xspace}

\newcommand{\raa}{\ensuremath{R_{AA}}\xspace}
\newcommand{\taa}{\ensuremath{T_{AA}}\xspace}

\newcommand{\eq}[1]{~\ref{#1}\xspace}
\newcommand{\fig}[1]{{Fig.~\ref{#1}}\xspace}
\newcommand{\tab}[1]{table~\ref{#1}\xspace}

\renewcommand{\eg}{\emph{e.\,g.}\xspace}%
\renewcommand{\ie}{\emph{i.\,e.}\xspace}%

\newcommand{\pp}{{\ensuremath{\text{pp}}}\xspace}
\newcommand{\ppbar}{{\ensuremath{\text{p}\overline{\text{p}}}}\xspace}
\newcommand{\PbPb}{\ensuremath{\text{PbPb}}\xspace}
\newcommand{\AuAu}{\ensuremath{\text{AuAu}}\xspace}

\newcommand{\sqrts}{\ensuremath{\sqrt{s}}\xspace}
\newcommand{\sqrtsnn}{\ensuremath{\sqrt{s_{NN}}}\xspace}

\newcommand{\pT}{\pt}
\newcommand{\mt}{\ensuremath{m_T}\xspace}

\newcommand{\mbinv} {\mbox{\ensuremath{\,\text{mb}^{-1}}}\xspace}

\newcommand{\pythia}{{\sc Pythia}\xspace}
\newcommand{\hydjet}{{\sc Hydjet}\xspace}
\newcommand{\cascade}{{\sc Cascade}\xspace}

\cmsNoteHeader{2012-216} % This is over-written in the CMS environment: useful as preprint no. for export versions
\title{Quarkonia and heavy-flavour production in CMS}

\author[llr]{Torsten Dahms,
  for the CMS collaboration}

\date{\today}

\abstract{
  The Compact Muon Solenoid (CMS) has measured numerous quarkonium
  states via their decays into \mumu pairs in \pp and \PbPb collisions
  at $\sqrtsnn = 2.76\TeV$. Quarkonia are especially relevant for
  studying the quark-gluon plasma since they are produced at early
  times and propagate through the medium, mapping its
  evolution. Non-prompt \Jpsi from b-hadron decays show a strong
  suppression in the transverse momentum range ($6.5<\pT<30\GeVc$)
  when compared to the yield in \pp collisions scaled by the number of
  inelastic nucleon-nucleon collisions. This suppression is related to
  the in-medium b-quark energy loss. In the same kinematic region, for
  prompt \Jpsi, a strong, centrality-dependent suppression is observed
  in \PbPb collisions. Such strong suppression at high \pT has
  previously not been observed at RHIC. At midrapidity ($|y|<1.6$) and
  the same \pT region, inclusive \psiP are even stronger suppressed
  than \Jpsi, whereas \psiP at forward rapidity ($1.6<|y|<2.4$) and
  lower \pT ($3<\pT<30\GeVc$) appear to be less suppressed than \Jpsi,
  however, with large uncertainties that prevent a
  conclusion. Furthermore, low-\pT \PgUb\ and \PgUc\ mesons are
  strongly suppressed in \PbPb collisions. The suppression of the
  \PgUa\ state is smaller than the suppression of the excited states
  and consistent with the suppression of the feed-down contribution
  only.
}

\hypersetup{%
pdfauthor={CMS Collaboration},%
pdftitle={Quarkonia and heavy-flavour production in CMS},%
pdfsubject={CMS},%
pdfkeywords={heavy-ion collisions quarkonium, charmonium, bottomonium,
  heavy flavour}}

\maketitle %maketitle comes after all the front information has been supplied

\section{Introduction}
\label{sec:intro}

The goal of the SPS, RHIC, and LHC heavy-ion programmes is to validate
the existence and study the properties of the quark-gluon plasma
(QGP), a state of deconfined quarks and gluons. One of its most
striking expected signatures is the suppression of quarkonium
states~\cite{Matsui:1986dk}, both of the charmonium (\Jpsi, \psiP,
$\chi_c$, etc.) and the bottomonium ($\PgU\text{(1S,\,2S,\,3S)}$,
$\chi_b$, etc.) families. This is thought to be a direct effect of
deconfinement, when the binding potential between the constituents of
a quarkonium state, a heavy quark and its antiquark, is screened by
the colour charges of the surrounding light quarks and gluons. The
suppression is predicted to occur above the critical temperature of
the medium ($T_c$) and depends on the \QQbar binding energy. Since the
\PgUa\ is the most tightly bound state among all quarkonia, it is
expected to be the one with the highest dissociation
temperature. Examples of dissociation temperatures are given in
Ref.~\cite{Mocsy:2007jz}: $T_{\text{dissoc}}\sim\!1\,T_c$, $1.2\,T_c,$
and $2\,T_c$ for the \PgUc, \PgUb, and \PgUa, respectively. Similarly,
in the charmonium family the dissociation temperatures are
$\leq1\,T_c$ and $1.2\,T_c$ for the \psiP and \Jpsi,
respectively. However, there are further possible changes to the
quarkonium production in heavy-ion collisions. On the one hand,
modifications to the parton distribution functions inside the nucleus
(shadowing) and other cold-nuclear-matter effects can reduce the
production of quarkonia without the presence of a
QGP~\cite{Vogt:2010aa,Zhao:2011cv}. On the other hand, the large
number of heavy quarks produced in heavy-ion collisions, in particular
at the energies accessible by the Large Hadron Collider (LHC), could
lead to an increased production of quarkonia via statistical
recombination~\cite{Zhao:2010nk,Andronic:2006ky,Capella:2007jv,Thews:2005vj,Yan:2006ve,Grandchamp:2005yw}.

In this proceedings, the CMS measurements of non-prompt and prompt
\Jpsi, inclusive \psiP, and the $\PgU\text{(1S,\,2S,\,3S)}$ mesons in
\PbPb collisions are discussed. The results are presented as nuclear
modifications factors (\raa), based on a comparison to the yield
measured in a \pp reference run at the same \sqrtsnn, scaled by the
number of binary collisions (\ncoll):
\begin{linenomath}
  \begin{align}
    \raa = \frac{\mathcal{L}_{\pp}}{\taa N_{\text{MB}}}\frac{N_{\PbPb} (\QQbar)}{N_{\pp} (\QQbar)}\cdot \frac{\varepsilon_{\pp}}{\varepsilon_{\PbPb}}\,.\notag
  \end{align}
\end{linenomath}
The measured yields in \PbPb ($N_{\PbPb}(\QQbar)$) and \pp collisions
($N_{\pp}(\QQbar)$) are corrected by their respective efficiencies
$\varepsilon_{\PbPb}$ and $\varepsilon_{\pp}$. $\mathcal{L}_{\pp}$ is
the integrated luminosity of the \pp dataset, $\taa$ is the nuclear
overlap function, which is equal to \ncoll divided by the elementary
nucleon-nucleon cross section, and $N_{\text{MB}}$ is the number of
minimum bias events in the \PbPb sample. While non-prompt and prompt
\Jpsi results are based on the 2010 dataset corresponding to an
integrated luminosity of $\mathcal{L}_{\text {int}} = 7.28\mubinv$,
the \psiP and \PgU\ results were obtained from the twenty times larger
2011 dataset with an integrated luminosity of $\mathcal{L}_{\text
  {int}} = 150\mubinv$. The \pp reference dataset used in all results
has an integrated luminosity of $\mathcal{L}_{\text {int}} =
231\nbinv$, which for hard-scattering processes is comparable in size
to the 2010 \PbPb sample ($7.28\mubinv \cdot 208^2 \approx
315\nbinv$). A more detailed discussion of the charmonium analyses
presented at this conference can be found in~\cite{Moon:2012a},
whereas details on the bottomonium analyses are discussed
in~\cite{Mironov:2012a}.

The central feature of CMS is a superconducting solenoid, of 6\,m
internal diameter, providing a field of 3.8\,T. Within the field
volume are the silicon pixel and strip tracker, the crystal
electromagnetic calorimeter (ECAL) and the brass/scintillator hadron
calorimeter (HCAL). Muons are measured in gas-ionization detectors
embedded in the steel return yoke. In addition to the barrel and
endcap detectors, CMS has extensive forward calorimetry. The muons are
measured in the pseudorapidity window $|\eta|< 2.4$, with detection
planes made of three technologies: Drift Tubes, Cathode Strip
Chambers, and Resistive Plate Chambers. Matching the muons to the
tracks measured in the silicon tracker results in a transverse
momentum resolution better than 1.5\% for \pT smaller than 100\GeVc. A
much more detailed description of CMS can be found
elsewhere~\cite{Adolphi:2008zzk}.

\begin{figure}[hbtp]
  \begin{center}
    \includegraphics[width=0.8\linewidth]{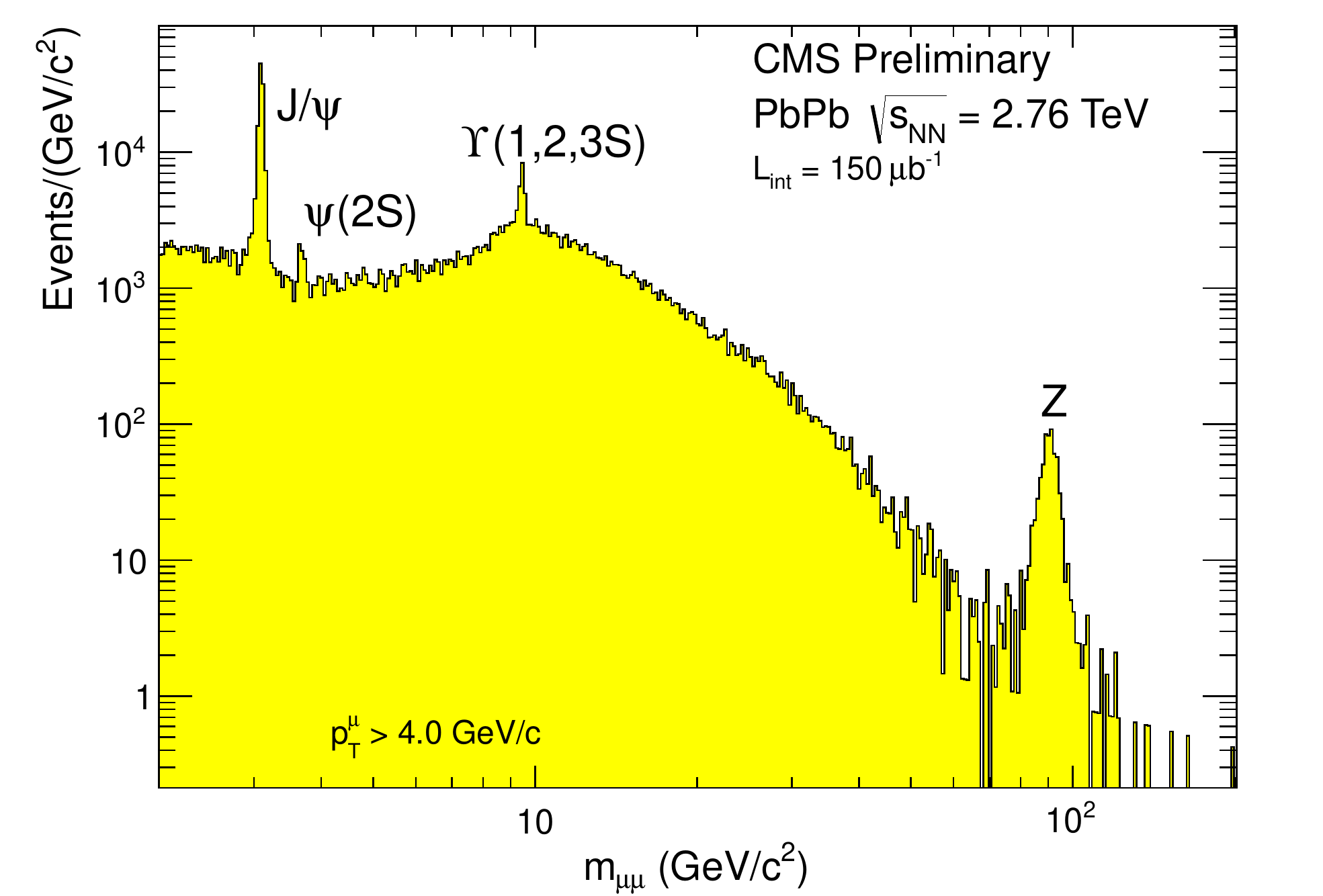}
    \caption{Invariant-mass spectrum of \mumu pairs measured in \PbPb
      collisions at $\sqrtsnn = 2.76\TeV$. A cut of $\pT>4\GeVc$ has
      been applied to both muons. Visible are the resonance peaks of
      the charmonium and bottomonium families, as well as the Z
      boson.}
    \label{fig:dimuonMass}
  \end{center}
\end{figure}

Figure~\ref{fig:dimuonMass} shows the invariant-mass spectrum of \mumu
pairs measured in \PbPb collisions at $\sqrtsnn = 2.76\TeV$. Both
muons of the pair are required to have a transverse momentum of at
least 4\GeVc. The spectrum ranges in mass from 2\GeVcc to 200\GeVcc
and demonstrates the excellent capabilities of the CMS detector to
measure dimuons over a broad kinematic range. Clearly reconstructed
are the resonance peaks of the charmonium and bottomonium families, as
well as the Z boson.

\section{Charmonia}
\label{sec:charm}

\begin{figure}[htp]
  \begin{center}
    \includegraphics[width=0.4\linewidth]{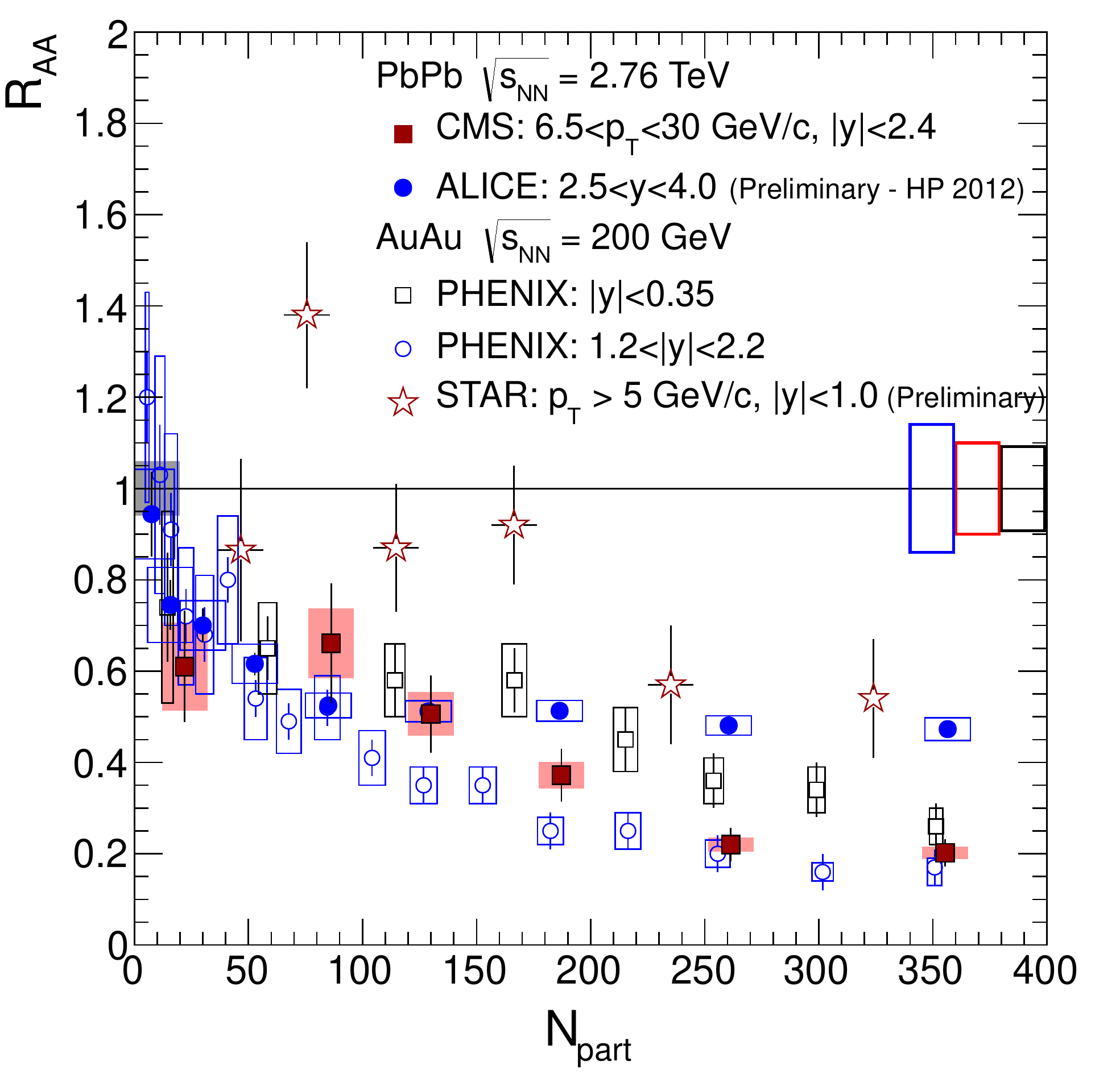}
    \includegraphics[width=0.4\linewidth]{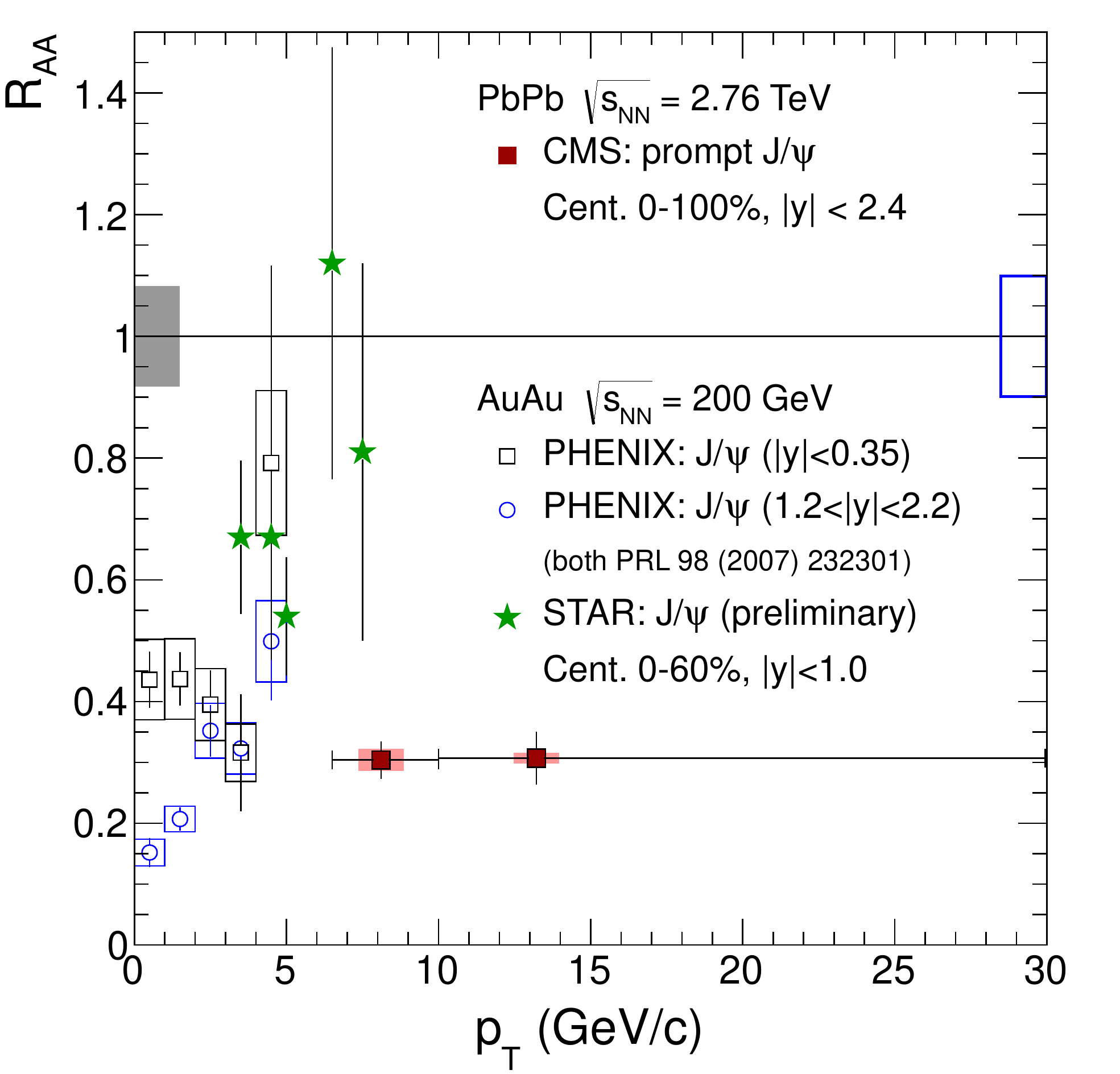}
    \caption{\Jpsi \raa as a function of centrality (left) and \pT
      (right). The prompt \Jpsi measurement of CMS (solid red squares)
      is compared to inclusive \Jpsi measurements by ALICE (solid blue
      circles), STAR (stars), and PHENIX at midrapidity (open black
      squares) and forward rapidity (open blue circles). Statistical
      (systematic) uncertainties are shown as bars (boxes). In case of
      the STAR results, statistical and systematic uncertainties are
      shown combined as bars. Global uncertainties from the \pp
      luminosity are shown as boxes at unity.}
    \label{fig:promptJpsi}
  \end{center}
\end{figure}

CMS has recently published a measurement of the nuclear modification
factor of prompt \Jpsi in \PbPb collisions at $\sqrtsnn = 2.76\TeV$ as
a function of \Jpsi \pT and rapidity, and event centrality, based on
an integrated luminosity of $\mathcal{L}_{\text {int}} =
7.28\mubinv$~\cite{Chatrchyan:2012np}. Non-prompt \Jpsi from b-hadron
decays and prompt \Jpsi have been separated in a two dimensional fit
to the invariant mass and the transverse distance between the
collision vertex and reconstructed secondary vertex of the \mumu
pair. The results show a strong, centrality-dependent suppression of
prompt \Jpsi in \PbPb collisions, compared to the yield in \pp
collisions scaled by the number of inelastic nucleon-nucleon
collisions (\ncoll). The centrality and \pT dependencies of the \raa
are shown in the left and right panel of \fig{fig:promptJpsi},
respectively. In the 10\% most central collisions, a nuclear
modification factor of $\raa =
0.20\pm0.03\,(\text{stat.})\pm0.01\,(\text{syst.})$ has been
measured. In peripheral \PbPb collisions (50--100\%), the suppression
is three times smaller than in the most central collisions. The prompt
\Jpsi \raa does not vary with \pT in the range $6.5<\pT<30\GeVc$. The
results are compared to other inclusive \Jpsi measurements at RHIC and
the LHC. The preliminary measurement of inclusive high-\pT \Jpsi at
RHIC by STAR~\cite{Tang:2011kr} shows a much smaller suppression, even
in the most central collisions at RHIC energies ($\sqrtsnn =
200\GeV$). In contrast, the low-\pT \Jpsi \raa measured at RHIC by the
PHENIX collaboration~\cite{Adare:2011yf} show a qualitatively similar
suppression pattern as the one measured by CMS. The ALICE \raa
measurement of low-\pT inclusive \Jpsi at forward rapidity
($2.5<y<4$)~\cite{Suire:2012gt} shows almost no centrality dependence
of the suppression, which is significantly smaller than the one of
high-\pT \Jpsi at the LHC. The difference between the strong
suppression of high-\pT \Jpsi measured by CMS and the \raa of low-\pT
\Jpsi measured by ALICE at forward rapidity, might be a sign of
recombination.

\begin{figure}[hbp]
  \begin{center}
    \includegraphics[width=0.8\linewidth]{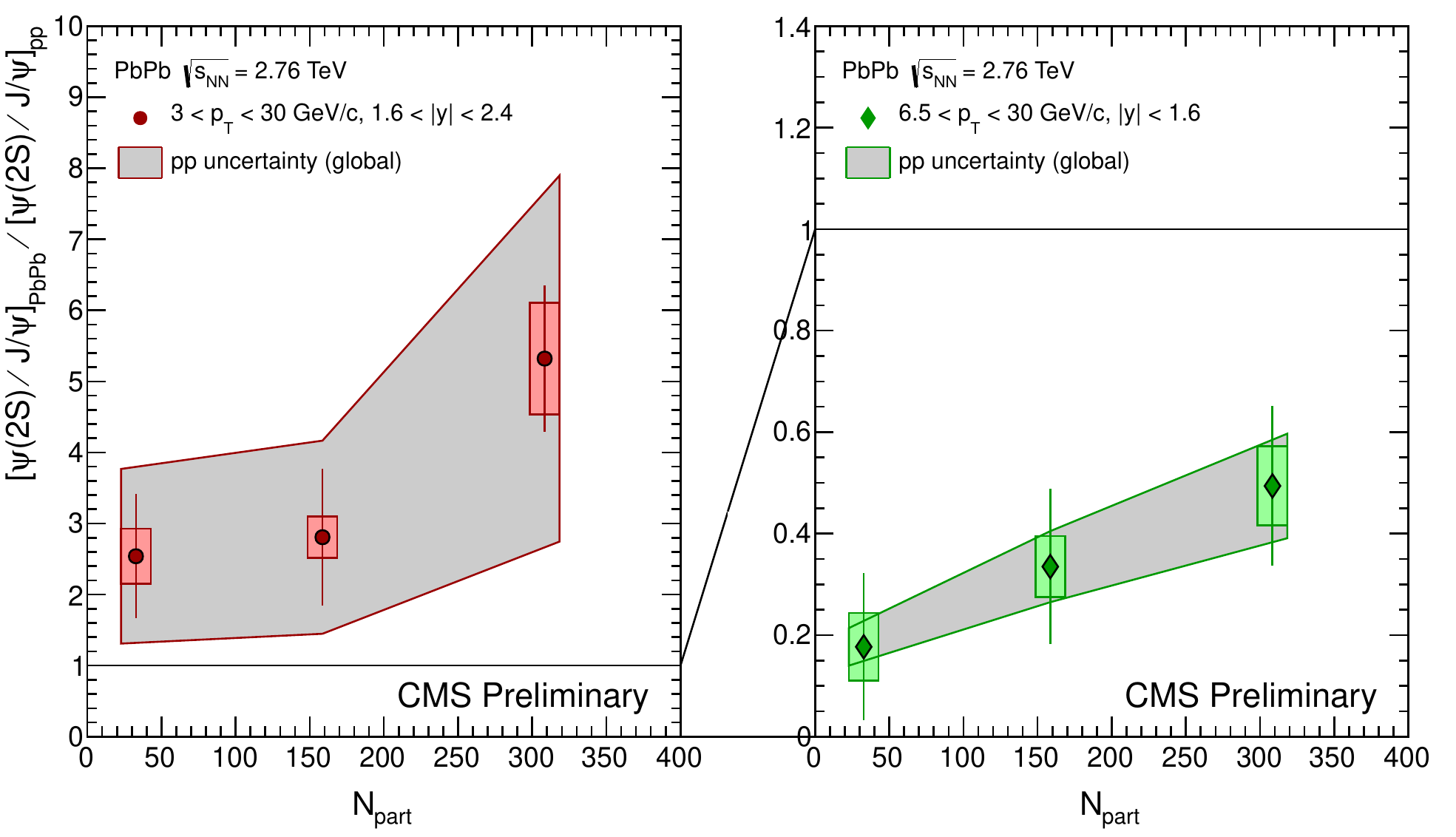}
    \caption{Centrality dependence of the double ratio \doubleRatioPsi
      at forward rapidity and low \pT ($1.6<|y|<2.4$ and
      $3<\pT<30\GeVc$) (left) and at midrapidity and high \pT
      ($|y|<1.6$ and $6.5<\pT<30\GeVc$) (right).}
    \label{fig:psiDoubleRatio}
  \end{center}
\end{figure}

More recently, CMS has also performed a preliminary measurement of the
relative suppression of inclusive \psiP mesons with respect to
inclusive \Jpsi in form of a double ratio
$(N_{\psiP}/N_{\Jpsi})_{\PbPb}/$\\$(N_{\psiP}/N_{\Jpsi})_{\pp}$
based on an integrated luminosity of $\mathcal{L}_{\text {int}} =
150\mubinv$ collected in 2011~\cite{CMS-PAS-HIN-12-007}. This double
ratio has been measured as a function of centrality at high \pT
($6.5<\pT<30\GeVc$) and midrapidity ($|y|<1.6$), as well as at lower
\pT ($3<\pT<30\GeVc$) and forward rapidity ($1.6<|y|<2.4$). For the
latter kinematic region a double ratio that increases with centrality
has been measured as shown in the left panel of
\fig{fig:psiDoubleRatio}, though with large uncertainties. In the 20\%
most central collisions, a double ratio of
$5.32\pm1.03\,\text{(stat.)}\pm0.79\,\text{(syst.)}
\pm\,2.58\,\text{(pp)}$ has been measured, which means that more \psiP
are produced compared to \Jpsi than in \pp collisions, again with
large uncertainties mostly originating from the limited size of the
\pp sample. However, in the right panel of \fig{fig:psiDoubleRatio} it
can be seen that at high \pT and midrapidity, the double ratio is
always less than unity, meaning that, in this kinematic range, \psiP
are more suppressed than \Jpsi. Within uncertainties, no centrality
dependence is observed. For the centrality integrated bins the double
ratio can be converted into an inclusive \psiP \raa, by simple
multiplication with the published inclusive \Jpsi
\raa~\cite{Chatrchyan:2012np}. For $6.5<\pT<30\GeVc$ and $|y|<1.6$ the
\psiP \raa is:
\begin{linenomath}
  \begin{align}
    \raa(\psiP) &= 0.11 \pm 0.03\,\text{(stat.)} \pm 0.02\,\text{(syst.)} \pm 0.02\,\text{(pp)}.\notag
  \end{align}
\end{linenomath}
For $3<\pT<30\GeVc$ and $1.6<|y|<2.4$ the \psiP \raa is:
\begin{linenomath}
  \begin{align}
    \raa(\psiP) &= 1.54 \pm 0.32\,\text{(stat.)} \pm 0.22\,\text{(syst.)} \pm 0.76\,\text{(pp)}.\notag
  \end{align}
\end{linenomath}

These results exhibit a clear \psiP suppression in the midrapidity and higher-\pT
region, while the \pp uncertainty is too large to draw a firm
conclusion in the forward rapidity lower-\pT region.

\section{Bottomonia}
\label{sec:bottom}

\begin{figure}[ht]
  \begin{center}
    \includegraphics[width=0.4\linewidth]{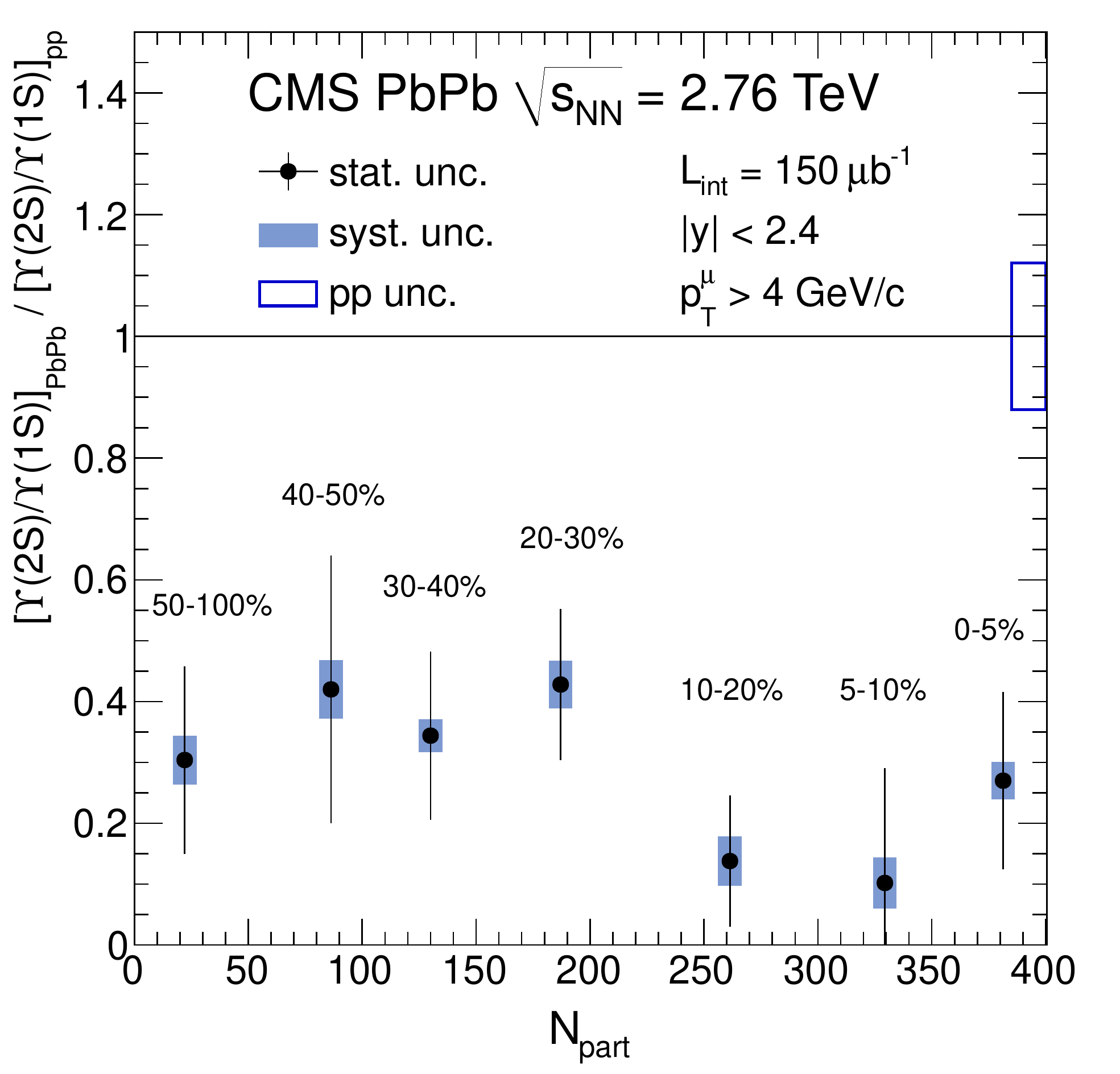}
    \includegraphics[width=0.4\linewidth]{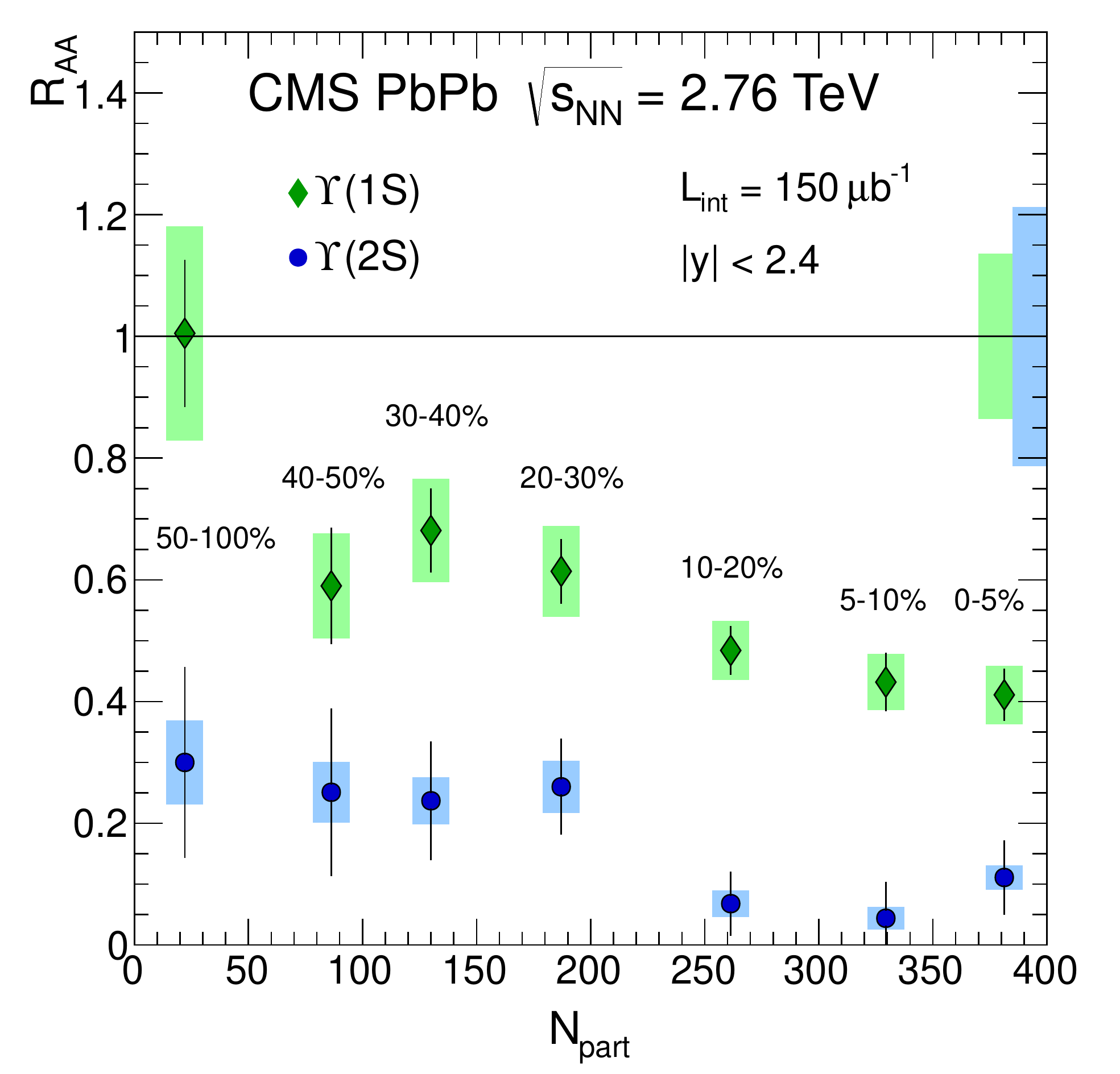}
    \caption{Centrality dependence of the double ratio \doubleRatioUps
      (left) and the nuclear modification factor of \PgUa\ and \PgUb\
      (right). Statistical (systematic) uncertainties are shown as
      bars (boxes). Global uncertainties from the \pp yields and, in
      case of the \raa, luminosity, are shown as boxes at unity.}
    \label{fig:upsilon}
  \end{center}
\end{figure}

CMS measured the \PgUa\ \raa in \PbPb collisions at $\sqrtsnn =
2.76\TeV$ for the first time in \cite{Chatrchyan:2012np} and the
suppression of the excited states relative to the \PgUa\ in
\cite{Chatrchyan:2011pe}. Based on an integrated luminosity of
$\mathcal{L}_{\text {int}} = 150\mubinv$ recorded in 2011, a more
detailed measurement has been made
possible~\cite{Chatrchyan:2012fr}. Centrality integrated double ratios
have been measured separately for the \PgUb\ and \PgUc\ states:
\begin{linenomath}
  \begin{align}
    \doubleRatioUps &= 0.21 \pm 0.07\,\text{(stat.)} \pm 0.02\,\text{(syst.)},\notag\\
    \doubleRatioUpsC &= 0.06 \pm 0.06\,\text{(stat.)} \pm 0.06\,\text{(syst.)}\quad(<0.17\text{ at 95\% CL}).\notag
  \end{align}
\end{linenomath}
The \PgUb\ double ratio has been measured as a function of centrality,
as shown in \fig{fig:upsilon}. Within uncertainties, no pronounced
centrality dependence is observed.  Furthermore, the nuclear
modification factors of all three states have been measured,
integrated over centrality:
\begin{linenomath}
  \begin{align}
    \raa(\PgUa) &= 0.56 \pm 0.08\,\text{(stat.)} \pm 0.07\,\text{(syst.)},\notag\\
    \raa(\PgUb) &= 0.12 \pm 0.04\,\text{(stat.)} \pm 0.02\,\text{(syst.)},\notag\\
    \raa(\PgUc) &= 0.03 \pm 0.04\,\text{(stat.)} \pm 0.01\,\text{(syst.)}\quad(<0.10\text{ at 95\% CL}).\notag\\
  \end{align}
\end{linenomath}
The centrality dependence of the nuclear modification factors of
\PgUa\ and \PgUb\ are shown in the right panel of \fig{fig:upsilon}.

The data show a clear ordering of the suppression with binding energy,
the least bound state being the most suppressed. The suppression of
the \PgUa\ state is consistent with no suppression of directly
produced \PgUa, but a suppression of feed-down contribution from
excited state decays only which is expected to contribute $\approx
50\%$ at high \pT~\cite{Affolder:1999wm}. However, the uncertainties
in the measurement of the feed-down contributions preclude
quantitative conclusions about the suppression of directly produced
\PgUa.

\section{Open heavy-flavour}
\label{sec:open}

\begin{figure}[ht]
  \begin{center}
    \includegraphics[width=0.4\linewidth]{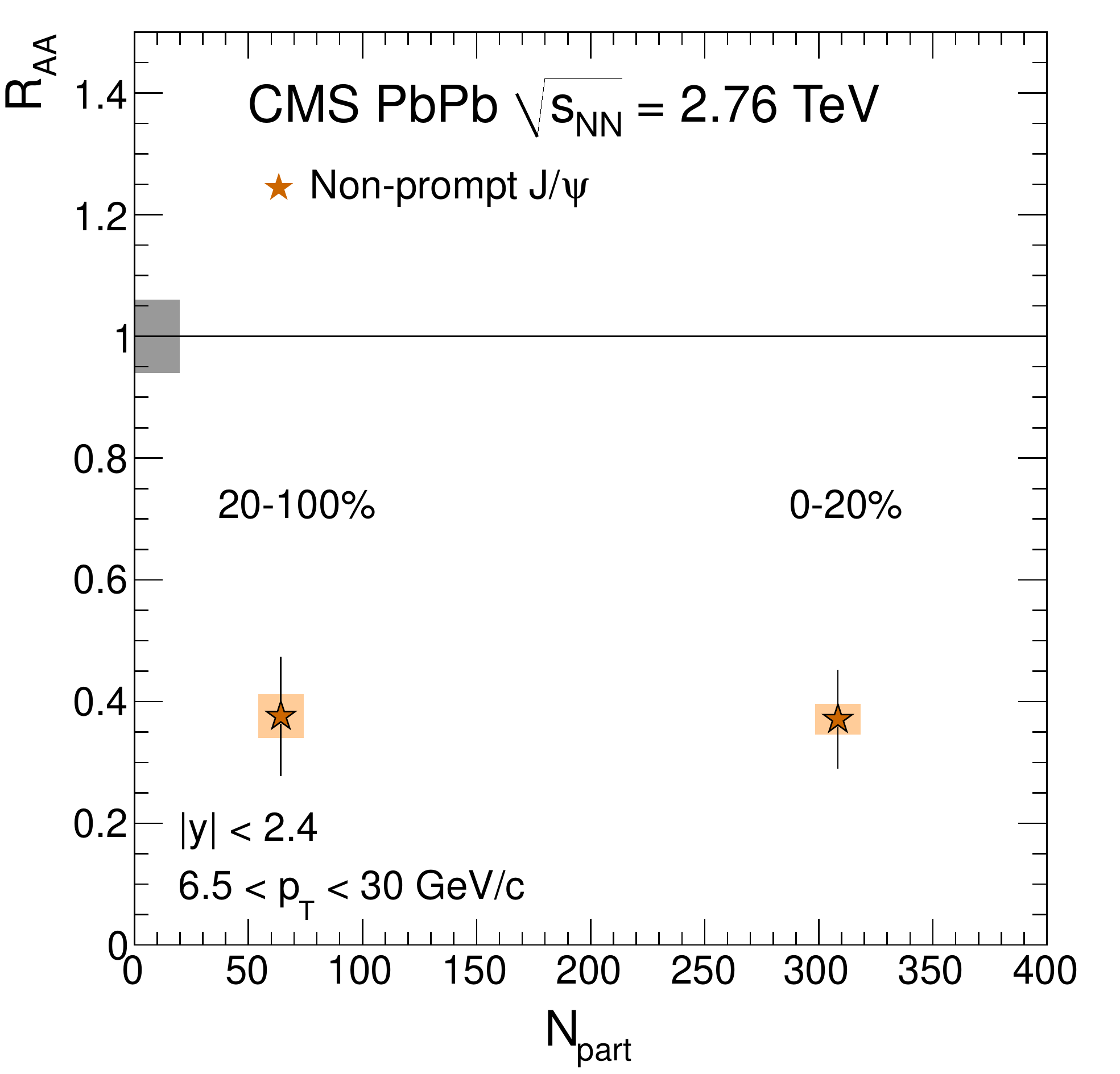}
    \includegraphics[width=0.4\linewidth]{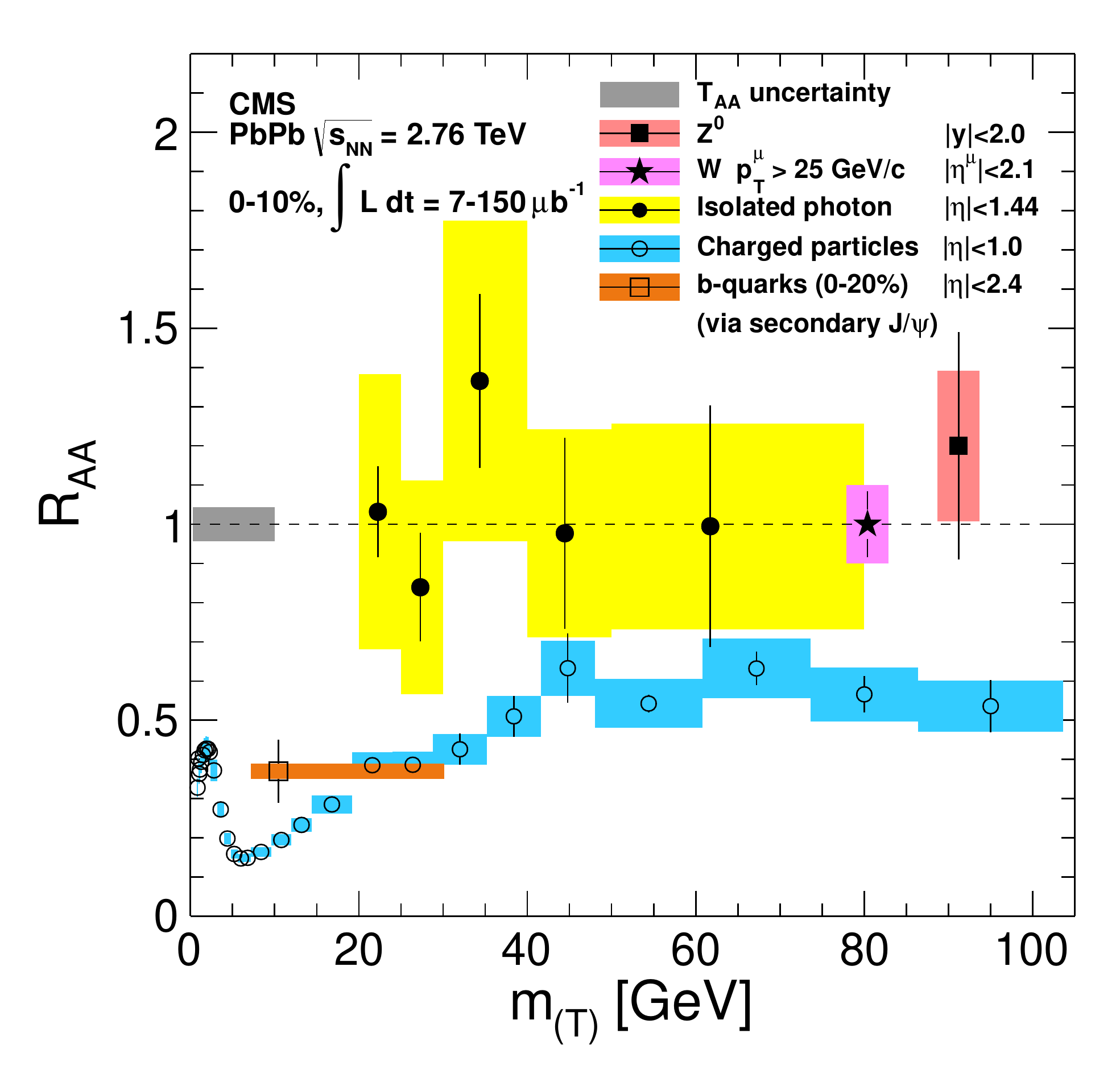}
    \caption{Centrality dependence of the nuclear modification factor
      of non-prompt \Jpsi from b-hadron decays (left) and \mt
      dependence of the light hadron \raa compared to non-prompt \Jpsi
      and electroweak bosons. Statistical (systematic) uncertainties
      are shown as bars (boxes). Global uncertainties from the \pp
      luminosity are shown as boxes at unity.}
    \label{fig:nonPromptJpsi}
  \end{center}
\end{figure}

As mentioned in section \ref{sec:charm}, CMS has separated prompt
\Jpsi and non-prompt \Jpsi from b-hadron decays in \PbPb
collisions~\cite{Chatrchyan:2012np}. The \raa of non-prompt \Jpsi with
$6.5<\pT<30\GeVc$ and $|y|<2.4$ is shown in two bins of centrality
(0--20\% and 20--100\%) in the left panel of
\fig{fig:nonPromptJpsi}. A clear suppression of \Jpsi from b-hadron
decays is observed. While the suppression is the same in the two
centrality bins, it is to be noted that the 20--100\% bin is very
broad and hard probes, such as b hadrons, will be produced
predominantly towards the more central edge of the bin. The right
panel of \fig{fig:nonPromptJpsi} shows a comparison of the non-prompt
\Jpsi \raa in the 0--20\% centrality bin compared to the charged
hadron \raa~\cite{CMS:2012aa}, which reflects the energy loss of light
quarks, and the \raa of electroweak
bosons~\cite{Chatrchyan:2011ua,Chatrchyan:2012nt,Chatrchyan:2012vq},
which do not interact strongly with the medium. More detailed
measurements of the \pT dependence of the non-prompt \Jpsi \raa are
necessary, before drawing conclusions on a possible mass hierarchy of
the in-medium energy loss of light and heavy quarks.

\section{Summary}
\label{sec:summary}

In summary, CMS has measured the nuclear modification factors of all
S-wave charmonium and bottomonium vector-meson states below twice the
D and B meson masses, respectively. For prompt \Jpsi, \PgUa, and
\PgUb\ the centrality dependence of the \raa has been
measured. Furthermore, the \raa as a function of \pT and rapidity has
been measured for prompt \Jpsi and \PgUa~\cite{Chatrchyan:2012np}. A
sequential melting of the \PgU\ states is observed. It is interesting
to note that also the suppression of prompt \Jpsi with
$6.5<\pT<30\GeVc$ and $|y|<2.4$, as well as inclusive \psiP with
$6.5<\pT<30\GeVc$ and $|y|<1.6$ follow the same ordering as a function
of the binding energy. However, the effect of the \pT and rapidity
cuts has to be evaluated. More detailed measurements of the \pT and
rapidity measurements are currently limited by the limited size of the
\pp reference dataset at \sqrts = 2.76\TeV. The same is true for the
measurement of the possible \psiP enhancement in the kinematic range
$3<\pT<30\GeVc$ and $1.6<|y|<2.4$, which is opposite to the expected
behaviour in the sequential melting scenario.

The in-medium energy loss of b quarks has been quantified via the
nuclear modification factor of non-prompt \Jpsi with $\pT>6.5\GeVc$,
which is comparable in magnitude to the one of light hadrons at high
\pT.

\section*{Acknowledgements}
\label{sec:ack}

TD received funding from the European Research Council under the FP7
Grant Agreement no. 259612.


\providecommand{\href}[2]{#2}\begingroup\raggedright\begin{thebibliography}{}%
\makeatletter
\providecommand{\hrefCMSnoop }[0]{\@secondoftwo}%
\makeatother
\providecommand{\doi}{\texttt{doi:}\begingroup \urlstyle{tt}\Url}

\end{thebibliography}\endgroup


\begin{thebibliography}{10}
\expandafter\ifx\csname url\endcsname\relax
  \def\url#1{\texttt{#1}}\fi
\expandafter\ifx\csname urlprefix\endcsname\relax\def\urlprefix{URL }\fi
\expandafter\ifx\csname href\endcsname\relax
  \def\href#1#2{#2} \def\path#1{#1}\fi

\bibitem{Matsui:1986dk}
T.~Matsui, H.~Satz, {\Jpsi suppression by quark-gluon plasma formation}, Phys.
  Lett. B 178 (1986) 416.
\newblock \href {http://dx.doi.org/10.1016/0370-2693(86)91404-8}
  {\path{doi:10.1016/0370-2693(86)91404-8}}.

\bibitem{Mocsy:2007jz}
{\'A}.~M{\'o}csy, P.~Petreczky, {Color screening melts quarkonium}, Phys. Rev.
  Lett. 99 (2007) 211602.
\newblock \href {http://arxiv.org/abs/0706.2183} {\path{arXiv:0706.2183}},
  \href {http://dx.doi.org/10.1103/PhysRevLett.99.211602}
  {\path{doi:10.1103/PhysRevLett.99.211602}}.

\bibitem{Vogt:2010aa}
R.~Vogt, {Cold Nuclear Matter Effects on \Jpsi and \PgU\ Production at energies
  available at the CERN Large Hadron Collider (LHC)}, Phys. Rev. C 81 (2010)
  044903.
\newblock \href {http://arxiv.org/abs/1003.3497} {\path{arXiv:1003.3497}},
  \href {http://dx.doi.org/10.1103/PhysRevC.81.044903}
  {\path{doi:10.1103/PhysRevC.81.044903}}.

\bibitem{Zhao:2011cv}
X.~Zhao, R.~Rapp, {Medium modifications and production of charmonia at LHC},
  Nucl. Phys. A 859 (2011) 114.
\newblock \href {http://arxiv.org/abs/1102.2194} {\path{arXiv:1102.2194}},
  \href {http://dx.doi.org/10.1016/j.nuclphysa.2011.05.001}
  {\path{doi:10.1016/j.nuclphysa.2011.05.001}}.

\bibitem{Zhao:2010nk}
X.~Zhao, R.~Rapp, {Charmonium in medium: From correlators to experiment}, Phys.
  Rev. C 82 (2010) 064905.
\newblock \href {http://arxiv.org/abs/1008.5328} {\path{arXiv:1008.5328}},
  \href {http://dx.doi.org/10.1103/PhysRevC.82.064905}
  {\path{doi:10.1103/PhysRevC.82.064905}}.

\bibitem{Andronic:2006ky}
A.~Andronic, P.~Braun-Munzinger, K.~Redlich, J.~Stachel, {Statistical
  hadronization of heavy quarks in ultra-relativistic nucleus--nucleus
  collisions}, Nucl. Phys. A 789 (2007) 334.
\newblock \href {http://arxiv.org/abs/nucl-th/0611023}
  {\path{arXiv:nucl-th/0611023}}, \href
  {http://dx.doi.org/10.1016/j.nuclphysa.2007.02.013}
  {\path{doi:10.1016/j.nuclphysa.2007.02.013}}.

\bibitem{Capella:2007jv}
A.~Capella, L.~Bravina, E.~Ferreiro, A.~Kaidalov, K.~Tywoniuk, E.~Zabrodin,
  {Charmonium dissociation and recombination at RHIC and LHC}, Eur. Phys. J. C
  58 (2008) 437.
\newblock \href {http://arxiv.org/abs/0712.4331} {\path{arXiv:0712.4331}},
  \href {http://dx.doi.org/10.1140/epjc/s10052-008-0772-6}
  {\path{doi:10.1140/epjc/s10052-008-0772-6}}.

\bibitem{Thews:2005vj}
R.~L. Thews, M.~L. Mangano, {Momentum spectra of charmonium produced in a
  quark-gluon plasma}, Phys. Rev. C 73 (2006) 014904.
\newblock \href {http://arxiv.org/abs/nucl-th/0505055}
  {\path{arXiv:nucl-th/0505055}}, \href
  {http://dx.doi.org/10.1103/PhysRevC.73.014904}
  {\path{doi:10.1103/PhysRevC.73.014904}}.

\bibitem{Yan:2006ve}
L.~Yan, P.~Zhuang, N.~Xu, {\Jpsi production in quark-gluon plasma}, Phys. Rev.
  Lett. 97 (2006) 232301.
\newblock \href {http://arxiv.org/abs/nucl-th/0608010}
  {\path{arXiv:nucl-th/0608010}}, \href
  {http://dx.doi.org/10.1103/PhysRevLett.97.232301}
  {\path{doi:10.1103/PhysRevLett.97.232301}}.

\bibitem{Grandchamp:2005yw}
L.~Grandchamp, S.~Lumpkins, D.~Sun, H.~van Hees, R.~Rapp, {Bottomonium
  production at \sqrtsnn = 200\GeV and \sqrtsnn = 5.5\TeV}, Phys. Rev. C 73
  (2006) 064906.
\newblock \href {http://arxiv.org/abs/hep-ph/0507314}
  {\path{arXiv:hep-ph/0507314}}, \href
  {http://dx.doi.org/10.1103/PhysRevC.73.064906}
  {\path{doi:10.1103/PhysRevC.73.064906}}.

\bibitem{Moon:2012a}
{D.~Moon, for the CMS collaboration}, {Measurement of charmonium production in
  \PbPb collisions at 2.76\TeV with CMS}, 2012, these proceedings.
\newblock \href {http://arxiv.org/abs/1209.1084} {\path{arXiv:1209.1084}}.

\bibitem{Mironov:2012a}
{C.~Mironov, for the CMS collaboration}, {Measurement of bottomonium production
  in \PbPb collisions at 2.76\TeV with CMS}, 2012, these proceedings.

\bibitem{Adolphi:2008zzk}
R.~Adolphi, et~al., {The CMS experiment at the CERN LHC}, JINST 3 (2008)
  S08004.
\newblock \href {http://dx.doi.org/10.1088/1748-0221/3/08/S08004}
  {\path{doi:10.1088/1748-0221/3/08/S08004}}.

\bibitem{Chatrchyan:2012np}
S.~Chatrchyan, et~al., {Suppression of non-prompt \Jpsi, prompt \Jpsi, and
  \PgUa\ in \PbPb collisions at \sqrtsnn = 2.76 TeV}, JHEP 05 (2012) 063.
\newblock \href {http://arxiv.org/abs/1201.5069} {\path{arXiv:1201.5069}},
  \href {http://dx.doi.org/10.1007/JHEP05(2012)063}
  {\path{doi:10.1007/JHEP05(2012)063}}.

\bibitem{Tang:2011kr}
{Z.~Tang, for the STAR collaboration}, {\Jpsi production and correlation in \pp
  and \AuAu collisions at STAR}, J. Phys. G 38 (2011) 124107.
\newblock \href {http://arxiv.org/abs/1107.0532} {\path{arXiv:1107.0532}},
  \href {http://dx.doi.org/10.1088/0954-3899/38/12/124107}
  {\path{doi:10.1088/0954-3899/38/12/124107}}.

\bibitem{Adare:2011yf}
A.~Adare, et~al., {\Jpsi suppression at forward rapidity in \AuAu collisions at
  \sqrtsnn = 200\GeV}, Phys. Rev. C 84 (2011) 054912.
\newblock \href {http://arxiv.org/abs/1103.6269} {\path{arXiv:1103.6269}},
  \href {http://dx.doi.org/10.1103/PhysRevC.84.054912}
  {\path{doi:10.1103/PhysRevC.84.054912}}.

\bibitem{Suire:2012gt}
{C.~Suire, for the ALICE collaboration}, {Charmonia production in ALICE}, 2012,
  these proceedings.
\newblock \href {http://arxiv.org/abs/1208.5601} {\path{arXiv:1208.5601}}.

\bibitem{CMS-PAS-HIN-12-007}
CMS, \href{http://cdsweb.cern.ch/record/1455477}{{Measurement of the \psiP
  meson in \PbPb collisions at \sqrtsnn = 2.76\TeV}}, CMS Physics Analysis
  Summary CMS-PAS-HIN-12-007 (2012).
\newline\urlprefix\url{http://cdsweb.cern.ch/record/1455477}

\bibitem{Chatrchyan:2011pe}
S.~Chatrchyan, et~al., {Indications of suppression of excited \PgU\ states in
  \PbPb collisions at \sqrtsnn = 2.76\TeV}, Phys. Rev. Lett. 107 (2011) 052302.
\newblock \href {http://arxiv.org/abs/1105.4894} {\path{arXiv:1105.4894}},
  \href {http://dx.doi.org/10.1103/PhysRevLett.107.052302}
  {\path{doi:10.1103/PhysRevLett.107.052302}}.

\bibitem{Chatrchyan:2012fr}
S.~Chatrchyan, et~al., {Observation of sequential \PgU\ suppression in \PbPb
  collisions}, submitted to Phys. Rev. Lett.\href
  {http://arxiv.org/abs/1208.2826} {\path{arXiv:1208.2826}}.

\bibitem{Affolder:1999wm}
A.~A. Affolder, et~al., {Production of \PgUa\ mesons from $\chi_b$ decays in
  \ppbar collisions at \sqrts = 1.8\TeV}, Phys. Rev. Lett. 84 (2000) 2094.
\newblock \href {http://arxiv.org/abs/hep-ex/9910025}
  {\path{arXiv:hep-ex/9910025}}, \href
  {http://dx.doi.org/10.1103/PhysRevLett.84.2094}
  {\path{doi:10.1103/PhysRevLett.84.2094}}.

\bibitem{CMS:2012aa}
S.~Chatrchyan, et~al., {Study of high-\pT charged particle suppression in \PbPb
  compared to \pp collisions at \sqrtsnn = 2.76\TeV}, Eur. Phys. J. C 72 (2012)
  1945.
\newblock \href {http://arxiv.org/abs/1202.2554} {\path{arXiv:1202.2554}},
  \href {http://dx.doi.org/10.1140/epjc/s10052-012-1945-x}
  {\path{doi:10.1140/epjc/s10052-012-1945-x}}.

\bibitem{Chatrchyan:2011ua}
S.~Chatrchyan, et~al., {Study of Z boson production in \PbPb collisions at
  \sqrtsnn = 2.76\TeV}, Phys.Rev.Lett. 106 (2011) 212301.
\newblock \href {http://arxiv.org/abs/1102.5435} {\path{arXiv:1102.5435}},
  \href {http://dx.doi.org/10.1103/PhysRevLett.106.212301}
  {\path{doi:10.1103/PhysRevLett.106.212301}}.

\bibitem{Chatrchyan:2012nt}
S.~Chatrchyan, et~al., {Study of W boson production in \PbPb and \pp collisions
  at \sqrtsnn = 2.76\TeV}, Phys. Lett. B 715 (2012) 66.
\newblock \href {http://arxiv.org/abs/1205.6334} {\path{arXiv:1205.6334}},
  \href {http://dx.doi.org/10.1016/j.physletb.2012.07.025}
  {\path{doi:10.1016/j.physletb.2012.07.025}}.

\bibitem{Chatrchyan:2012vq}
S.~Chatrchyan, et~al., {Measurement of isolated photon production in \pp and
  \PbPb collisions at \sqrtsnn = 2.76\TeV}, Phys. Lett. B 710 (2012) 256.
\newblock \href {http://arxiv.org/abs/1201.3093} {\path{arXiv:1201.3093}},
  \href {http://dx.doi.org/10.1016/j.physletb.2012.02.077}
  {\path{doi:10.1016/j.physletb.2012.02.077}}.

\end{thebibliography}
\end{document}